\documentstyle[preprint,prc,aps,epsfig]{revtex}
\def\shiftleft#1{#1\llap{#1\hskip 0.04em}}
\def\shiftdown#1{#1\llap{\lower.04ex\hbox{#1}}}
\def\thick#1{\shiftdown{\shiftleft{#1}}}
\def\b#1{\thick{\hbox{$#1$}}}
\begin{document}
\title{To the nature of intermediate- and short-range nucleon-nucleon
interaction}
\author{Vladimir I. Kukulin, I. T. Obukhovsky, V. N. Pomerantsev }
\address{Institute of Nuclear Physics, Moscow State
University, 119899 Moscow, Russia}

\maketitle

\begin{abstract} Instead of the Yukawa mechanism for intermediate- and
short-range interaction some new approach based on formation of the symmetric
six-quark bag in the state   $|(0s)^6[6]_X,L=0\rangle$ dressed due to strong
coupling to $\pi$, $\sigma$ and  $\rho$ fields are suggested. This new
mechanism offers both a strong intermediate-range attraction which replaces
the  effective $\sigma$-exchange (or excitation of two isobars in the
intermediate state) in traditional force models and also short-range
repulsion. Simple illustrative  model is developed which demonstrates
clearly
how well the suggested new mechanism can reproduce $NN$ data.
\end{abstract}

It was found in recent years that the traditional models for $NN$
forces, based on the Yukawa concept of one- or two-meson exchanges
between free nucleons~\cite{Mach87} even at the quark level
lead to very numerous
disagreements with newest precise experimental data for few-nucleon
observables (especially for spin-polarized
particles)~\cite{Kuk99,FB98,Fost99,Pla94}. There are also
numerous inner inconsistencies and disagreements between the
traditional $NN$ force models and predictions of fundamental theories
for meson-baryon interaction~\cite{Pla94} (e.g. for meson-nucleon
form factors). All these
disagreements  stimulate strongly the new attempts to develop
alternative force models based either on chiral perturbation theory
or a new quark-meson models.
Our recent studies in the field~\cite{Kuk99,Kus91,PRC99}
have led us to a principally new mechanism for intermediate- and
short-range $NN$ forces. This mechanism can also shed some
light to the puzzles in baryon spectroscopy (e.g. normal ordering in
$\Lambda$-sector and inverse ordering in nucleon sector for excited
negative and positive parity states).
The new model is based on the important observation~\cite{Kus91,PRC99}
that two possible
six-quark space symmetries in even $NN$ partial waves, viz.
$|s^6[6]L=0\rangle$ and $|s^4p^2[42]L=0,2\rangle$  which are allowed for $NN$ system
in $s$- and $d$-partial waves corresponds to the
states of different nature. The former states have almost the same
projections into various baryon-baryon (i.e. $NN$, $\Delta\Delta$, $CC$) channels and
thus correspond to bag-like
intermediate states while the coherent superposition of the
states of second type are projected mainly
into $NN$ channel with a large weight and thus can be presented as
clusterised $NN$ states with {\em nodal} $NN$ relative motion
wave functions~\cite{Kus91,PRC99} which
are similar to those derived from our previous Moscow potential
model~\cite{PRC99}.

 In the present work we develop this picture much
further on the quark-meson microscopic basis and derive the
microscopic $NN$ transition amplitudes through six-quark $+2\pi$
intermediate states in $s$-channel (see Fig.~1).
The transition is accompanied by a virtual emission and subsequent
absorption of two tightly correlated pions by diquark pairs or, alternatively,
by two $1p$-shell quarks when they jump from $1p$- to the $0s$-shell orbit or
vice versa. These
two pions can form both the scalar $\sigma$ and vector $\rho$ mesons
which surround the symmetric six-quark bag. Thus we adopt the $s$-channel
quark-meson   intermediate states as shown in Fig.1, the transition
amplitude being determined by $s$-channel   singularities in sharp contrast
to the Yukawa mechanism driven by   $t$-channel meson exchange.
This $s$-channel mechanism, being combined with an additional orthogonality
requirement~\cite{PRC99} (see below), can describe both the short-range repulsion and the medium
range attraction and can replace the $t$-channel exchange by $\sigma$- and
$\omega$-mesons in the conventional Yukawa-type picture of the $NN$ force\footnote
{Surely together with
this specific six-quark mechanism we should take into consideration also the
traditional Yukawa   mechanism for $\pi$-, $2\pi$- and $\rho$- (but not
$\sigma$-) meson exchanges between   isolated nucleons. However these
meson-exchange contributions are essential   only at the separations beyond
the intermediate six-quark bag ($>1$~fm) or in high   partial waves ($L>3$). In the
lowest partial waves, the intermediate dressed six-quark bag gives a
dominating contribution for the total $NN$ interaction.}.

In multiquark systems or in high density nuclear matter
some phase transition may happen when the quark density or the
temperature of the system is increased~\cite{Kun94}. This phase transition leads to a
partial restoration of the  broken  chiral symmetry and thus to the reduction of the
$\sigma$-meson and constituent-quark masses~\cite{Kun94}.
The significant reduction of $\rho$-meson mass in the nuclear
matter has been predicted also by the so called Brown-Rho
scaling. The most probable consequence of the restoration
should be strengthening of the sigma-meson field in the $NN$ overlap region.
This could be modeled by "dressing" of the most compact six-quark
configurations $\vert s^6[6]_XL=0\rangle$ and $\vert s^5p[51]_XL=1\rangle$
inside the $NN$ overlap region with an effective sigma-meson field. The
"$\sigma$" or a similar "scalar-isoscalar meson" is assumed to exist only in
a high density environment and  not in the vacuum, contrary to the $\pi$ and
$\rho$ mesons~\cite{Ku2000}. Thus the scalar- and vector-meson clouds will stabilize the
multi-quark bag due to a partial chiral symmetry restoration effect in the
dense multi-quark system and thus enhance all the contributions of such a
type. The picture of $NN$ interaction emerged from the model  can be
referred to as the "dressed" $6q$ bag (DB) model (see  Fig.~1).

Starting from this quark-hadron picture we assume that the total wave
function of the system $\Psi^L_{\rm total}$ at short $NN$ distances
(in lower partial waves L=0, 1)
consists of two parts with different nature: the "proper $NN$ component",
the quark-cluster part of which,
$\Psi^L_{NN}(6q)$, is dominated by the excited six-quark
configurations $s^4p^2$ at L=0 (or the $s^3p^3$ at L=1), and the
"proper dressed-bag component", the quark part of which,
$\Psi^L_0(6q)$, is dominated by the compact configurations $s^6$ (or the $s^5p$
at L=1) with a maximal overlap of all six quarks. The
bag-like configurations $s^6$ and $s^5p$ which are dressed by an enhanced
$\sigma$ field, viz. $\Psi^L_{DB}=\vert { }
6q+\sigma(\pi\pi)\rangle$   plays the same role in the hadronic sector of
our model as the $\Delta\Delta+\pi\pi$ intermediate state in the standard
(hadron) models of the $NN$ interaction~\cite{Mach87}.
However, the dressed bag component $\vert
6q+\sigma(\pi\pi)\rangle$ has a much more extended physical content than the
$\Delta\Delta+\pi\pi$ intermediate state in the traditional $NN$-models as:
({\it i}) the six-quark part of the DB implies a coherent sum over all the
possible baryon-baryon pairs in the cluster decomposition
$3q+3q$~\cite{Kus91,Ku2000}; ({\it ii}) the $\sigma$-meson (or $\pi+\pi$) part of the DB is
probably enhanced due to the (partial) chiral symmetry restoration which
implies the $\sigma$-meson and constituent quark masses to be noticeably
reduced~\cite{Kun94}\footnote{
The "compact" configurations $s^6$ and $s^5p$ are
usually included in the resonating-group-method (RGM) calculations for the
$NN$ system but without the strong $\sigma$ and $\rho$-meson fields
they play quite a passive role
providing only "dying out" of the $NN$ wave function at short range as a
result of the strong $NN$ repulsion in these quark states~\cite{Stancu}.}.

Thus we can treat the DB states as a new component in the Fock space or a
new (closed at $E<E_0\sim 600$~MeV) channel in the coupled channel approach to the $NN$ scattering
and write the total $NN$ wave function in the
form
\begin{equation}
\Psi^{L}_{\rm total}=\left(\begin{tabular}{c} $\Psi^{L}_{NN}(6q)$\\
$\Psi^{L}_{DB}(6q+\sigma)$\end{tabular}\right),\qquad
\Psi^{L}_{DB}(6q+\sigma)=\Psi^{L}_0(6q)\otimes\sigma(\pi\pi), \label{Fock}
\end{equation}
where the "proper" $NN$ wave function $\Psi^{L}_{NN}(6q)$ is orthogonal
to the six-quark part of the DB component $\Psi^{L}_0(6q)$~\cite{PRC99},
i.e. $\langle\Psi^{L}_{NN}(6q)\vert\Psi^{L}_0(6q)\rangle=0$, at L=0,1.

The transition amplitude
from the initial $s^4p^2(L=0,\,2)$ (or $s^3p^3$ at $L=1,\,3$)
six-quark configurations (the coherent superposition of those corresponds to
the proper $NN$ channel) to the intermediate ones $s^6(L=0)+(\pi\pi)$ (or
$s^5p + (\pi\pi)$ at $L=1$)  is accompanied with an emission of the
S-wave correlated $\pi+\pi$ pair, the both pions being  created in the $s$-wave due to
conservation of parity and angular momentum.

The intermediate six-quark configuration $s^5p[51]_X$ (denoted by vertical
dashed lines in the Fig.~1) have fixed quantum numbers which are determined by
the initial ($NN$) and intermediate ("dressed" bag $6q+\sigma$) states. The
second (after the first pion emission) state in the channel
$ST=01,\,\,J^P=0^+$ has quantum numbers of the so-called
$d^{\prime}$-dibaryon
$d^{\prime} =|(0s)^5(1p)[51]_XL\!=\!1,[321]_{CS}(ST=10)J^P=0^-\rangle$
(see, e.g.~\cite{Obu99}). In the channel
$ST=10,\,\,J^P=1^+$ the transition goes via another intermediate state
$d^{\prime\prime}=|(0s)^5(1p)[51]_XL\!=\!1\,[2^21^2]_{CS}(ST=01)
J^P=1^-\rangle$, which is a partner of the $d^{\prime}$ by
$S\leftrightarrow T$ interchanged.
Both intermediate configurations $d^{\prime}$
and $d^{\prime\prime}$ cannot decay into the two-nucleon channel as their
quantum numbers do not satisfy the Pauli exclusion principle for two
nucleons~\cite{Obu99}.

The transition amplitude is calculated here in the framework of the well
known quark-pair-creation model (QPCM) \cite{Micu69} (see also
\cite{Obu99}) and the transition operator for the emission of
the pion $\pi^{\lambda}$ ($\lambda=0,\pm$) by a single (e.g., the j-th)
quark in a six-quark system  $H^{(j)}_{\lambda}({\b k}_j)$ is taken in
the form proposed earlier~\cite{Obu99}.
The $\pi
+\pi \to \sigma$ transition amplitude is determined~\cite{Guts94} to be
proportional to the overlap of the two-pion and the $\sigma$-meson wave
functions: $\langle \pi({\b k})\pi({\b
k^{\prime}})|H_{\pi\pi\sigma}|\,\sigma\rangle
=g_{\pi\pi\sigma}F_{\pi\pi\sigma}(({\b k}-{\b k^{\prime}})^2)$, $F({\b
k}^2)=\exp(-\frac{1}{2}k^2b_{\sigma}^2)$.
In the limit of a point-like pion
the operator $H^{(6)}_{\lambda}$
goes to the standard pseudo-vector (PV) quark-pion coupling. Thus the
phenomenological coupling constant of the QPCM is normalized to the standard
PV $\pi qq$ coupling constant $f_{\pi qq}=\frac{3}{5}f_{\pi NN}$.

The amplitude $NN^{L=0,2}(s^4p^2) \to d'(d'')+\pi \to 6q(s^6)+\sigma$ can be
expressed through a matrix element of a transition operator $\Omega_{NN\to
d_0+\sigma}$:
$
A^{L=0(2)}_{NN\to d_0+\sigma}(E;{\b k})=\int
d^3r\,\Psi_{NN}^{L=0(2)}(E;{\b r})\, \Omega_{NN\to d_0+\sigma}(E;{\b
r},{\b k})\, $,
where $\Psi_{NN}^{L=0(2)}(E;{\b r})$ is the proper $NN$ wave function in the
sense of Eq.~(\ref{Fock}), $E=2m_N+p_N^2/m_N$ and
the plane-wave approximation  is used for the intermediate DB state
$d_0+\sigma$.
The operator $\Omega_{NN\to d_0+\sigma}$ incorporates the contribution of
the six-quark state (in the left hand part of the graph of FIG.~1) projected onto
two-nucleon clusters of the initial state and
can be written as an integral
of the elementary six-quark transition amplitude over both inner coordinates
of quark clusters (viz. N(123), N(456), $\pi$ and $\sigma$) and the pion
momenta ${\b k}_5$ and ${\b k}_6$:
\begin{eqnarray}
&\Omega_{NN\to d_0+\sigma}(E;{\b r},{\b k})= 15 { \displaystyle\int
d^3k_5\,\int d^3k_6\, \delta({\b k}_5+{\b k}_6-{\b k})}&\nonumber\\
&{\times}\sum\limits_{\lambda}\frac{(-1)^{1-\lambda}}{\sqrt{3}}\,\,
\frac{\sqrt{10}\,\langle NN|H_{\lambda}^{(6)}({\b k}_6)
|\,d^{\prime}(d^{\prime\prime})\rangle\,\langle
d^{\prime}(d^{\prime\prime})| H_{-\lambda}^{(5)}({\b k}_5)|\,d_0\rangle
\,g_{\pi\pi\sigma}F_{\pi\pi\sigma}(({\b k}_5-{\b k}_6)^2)}
{\left[m_{d^{\prime}}+\frac{k_6^2}{2m_{d^{\prime}}}+\omega_{\pi}(k_6)-
E\right]\left[m_{d_0}+\frac{k^2}{2m_{d_0}}+\omega_{\pi}(k_5)+
\omega_{\pi}(k_6)-E\right]}&
\label{om}
\end{eqnarray}
The numerical factor 15 in front of the integral counts the number of
$qq$ pairs in the six-quark system.

 All the matrix elements of interest are calculated using the f.p.c.
technique \cite{Kus91,Harv81} and are reduced to a product of the
vertex constant $vf_{\pi AB}$ (in the QPCM $v=-i(f_{\pi qq}/m_{\pi}) \left
((2\pi)^{3}2\omega_{\pi}\right)^{-1/2}$), form factor $F_{\pi{}AB}(k_i)$ and
a kinematical factor $\omega_{\pi}(k_i)/m_qb$~\cite{Obu99}):
\begin{equation}
\langle d_f^{L=0(2)}|H^{(6)}_{\lambda}({\b k}_6)|d^{\prime}\rangle =
v\,\frac{\omega_{\pi}(k_6)}{m_qb}f_{\pi d_fd^{\prime}}^{L} F^{L}_{\pi
d_fd^{\prime}}(k_6^2)\,\Sigma^{d_fd^{\prime}}\, T_{-\lambda}^{d_fd^{\prime}}
\label{fF}
\end{equation}
(and a similar expression for $\langle d^{\prime}|H^{(5)}_{\lambda}({\b
k}_5)|d_0\rangle$). In Eq.~(\ref{fF}) the $\Sigma^{d_fd^{\prime}}$ and
$T_{-\lambda}^{d_fd^{\prime}}$ are transition operators in the space of
total spin and isospin of the six-quark states $d_f$ and $d^{\prime}$. The
transition form factors $F^L$ depend on the angular momentum
L=0(2) of the
initial state: $F^{L}_{\pi d_fd^{\prime}}(k_6^2)= (1+a_{\scriptstyle L}
 \frac{5k_6^2 b_{\scriptstyle N}^2}{24}) \exp (-5k_6^2 b_{\scriptstyle N}^2/24)$,
$F_{\pi d_0d^{\prime}}(k_5^2)=\exp (-5k_5^2b_{\scriptstyle N}^2/24)$, with
 $a_{\scriptstyle L=0}=\frac{1}{3}$
and $a_{\scriptstyle L=2}=-\frac{2}{3}$. As a result one gets quite a simple
expression for the transition operator (\ref{om}) in the case of S and D
partial waves in the initial $NN$ channel:
\begin{equation}
\Omega_{NN\to d_0+\sigma}^{L}(E;{\b r},{\b k})= g_{\scriptstyle L}
e^{-5k^2b_{\scriptscriptstyle N}^2/48}\, D^L(E,k)\varphi_{2L}
({\b r}), \; L=0,2,
\label{omL}
\end{equation}
where $\varphi_{2L}
({\b r})$ is the h.o. wavefunction and the value $g_{L}$ means an effective strength constant of
the transition $N+N\to d_0+\sigma$ from initial (cluster-like) $NN$ state to
the intermediate "dressed" bag configuration, which takes the form:
\begin{equation}
g_{L}= \frac{f_{\pi
qq}^2}{m_{\pi}^2}\frac{g_{\pi\pi\sigma}}{m_q^2b_{\scriptstyle N}^2}\,C_L,
\quad \mbox{with } C_{L}=15f_{\pi d_0d^{\prime}}\sum\limits_f f_{\pi d_fd^{\prime}}^{L}
\Gamma^L_{d_f}U^{NN}_f =\frac{1}{144}{\scriptstyle\times}
\left\{\begin{array}{cr}-\frac{617}{2020\sqrt{5}},&L=0\\ \frac{11}{81},&L=2.
\end{array}\right..
\label{gL}
\end{equation}
The factor in $\Gamma^L_{d_f}$ in Eq.~(\ref{gL}) is a coordinate part of the
f.p.c. of the translationally-invariant shell model (TISM), while the
$U^{NN}_f$ is the respective CST part of it~\cite{Kus91,Obu99}. The function
$D^L(E,k)$ in Eq.~(\ref{omL}) corresponds to the loop integration in
Eq.~(\ref{om}).
Thus the calculation of the multi-loop diagram in Fig.~1 results in a
separable amplitude of the $NN$ interaction with left- and right-hand-side
vertices being expressed in the form
(\ref{omL})  while the loop
integral over the intermediate
$\vert s^6+\sigma\rangle$ state is expressed through the function
(a generalized propagator of the dressed bag):
\begin{equation}
G_{LL^{\prime}}(E)= \int \frac{k^2dk}{2{\cal E}(k)} \frac{\exp\left (
-\frac{5}{24}k^2b_{\scriptstyle N}^2\right )
D^L(E,k) D^{L'}(E,k)} {E-{\cal E}(k)},\;
{\cal E}(k)=m_{\sigma}+m_{d_0}+\frac{k^2}{2m_{\sigma
d_0}}, \label{am10}
\end{equation}
where the $m_{\sigma d_0}=m_{\sigma}m_{d_0}/(m_{\sigma}+m_{d_0})$ is the reduced mass
in the DB state.
In accordance with this, the contribution of the mechanism of Fig.~1
to the $NN$ interaction in the $S$ and $D$ partial waves
can be expressed through the matrix element:
$
A^{L^{\prime}L}_{NN\to d_0+\sigma\to NN}= \int
d^3r^{\prime}d^3r{\Psi^{L^{\prime}}_{NN}}^*(E;{\b r^{\prime}})
V^{L^{\prime}L}_E({\b r^{\prime}},{\b r})\Psi^L_{NN}(E; {\b r})\,
$,
where $V^{L'L}_E({\b r'},{\b r})$ is a separable potential
matrix of the form
\begin{equation}
V^{L'L}_E({\b r'},{\b r})
=\left(\begin{array}{cc}
g_0^2G_{00}(E)\phi_{2s}({\b r'})\phi_{2s}({\b r})
&g_0g_2G_{02}(E)\phi_{2s}({\b r'})\phi_{2d}({\b r})\\[5pt]
g_2g_0G_{20}(E)\phi_{2d}({\b r'})\phi_{2s}({\b r})
 &g_2^2G_{22}(E)\phi_{2d}({\b r'})\phi_{2d}({\b r})
\end{array}
\right). \label{sp}
\end{equation}
The interaction operator (\ref{sp}) mixes $S$- and $D$-partial waves in the triplet
$NN$ channel and thus it leads to a specific tensor mixing with the range
$\sim 1\,\,fm$ (about that of the intermediate DB state).
Thus the proposed new mechanism
results in a specific matrix separable form of the $NN$ interaction with
nodal (in S- in P-partial waves) form factors\footnote{This nodal character
of the form factors makes it possible to explain within this mechanism the
origin of the $NN$ repulsive core by the nodes in transition form factors.}
and a specific tensor mixing of new type \cite{KukMac}.

In the case of (partial) restoration of the chiral symmetry inside the
(compact) symmetric six-quark bag the effective $\sigma$-meson mass and
width should be much lower than its values accepted for the conventional OBE
models~\cite{Mach87,Kuk99}.
Then the position of the branch point $E_0=m_{d_0}+m_{\sigma}$ of the
function $G_{LL^{\prime}}(E)$ in Eq.~(\ref{sp}) will shift to lower energies and the
contribution of this (attractive) mechanism to the low-energy $NN$
interaction  becomes more important.
In other words, instead of the (artificial) increase of the cut-off
parameters in the $\pi NN\, (\sigma NN,\,\,\rho NN, \dots,\,\, etc.)$ form
factors in traditional OBE models~\cite{Mach87} we adopt the (natural) decrease of the denominator in
Eq.~(\ref{am10}). Furthermore the complicated energy dependence dictated by
Eq.(\ref{am10}) may be well approximated by a pole term $\sim
(E-\bar{E}_0)^{-1}$ with the effective pole position $\bar{E}_0$ either
calculated or simply fitted to the $NN$ phase shifts.

To illustrate the proposed new mechanism we built a simple model which
includes the main features of the suggested mechanism for
$NN$-interaction~\cite{Ku2000}. The model includes only a few parameters
for coupled partial waves and can fit the $NN$ phase shifts on the broad
energy range $0\div 600$~MeV, or even until 1200~MeV, using a {\em soft}
cut-off parameter $\Lambda_{\pi NN}\sim 0.6$~CeV
(in sharp contrast to any OBE model)
which is required by all microscopic models for meson-baryon coupling. The
model interaction consists of three terms: $ V_{NN}=V_{\rm orth}+V_{NqN} +V_{\rm OPE}$
with $V_{\rm orth}=\lambda_0 |\varphi_0\rangle \langle \varphi_0 |$,
$(\lambda_0 \to \infty)$,  providing the condition of orthogonality  between the proper
$NN$ channel and  the six-quark part of the intermediate  bag in $S$- and
$P$-waves. The one-pion-exchange potential  $V_{\rm OPE}$ is taken here with {\em soft}
dipole cut-off, while the separable  term $V_{NqN}$  for the single channel
case takes the form:
$V_{NqN}=E_0^2/(E-E_0) \lambda|\varphi\rangle \langle \varphi|$,
and  for coupled  channels it is a $(2{\times}
2)$-matrix:
\begin{equation}
V_{NqN}=\frac{E_0^2}{E-E_0}\left (
\begin{array}{cc}
\lambda_{11}|\varphi_1\rangle \langle \varphi_1| \qquad &
\lambda_{12}|\varphi_1\rangle \langle \varphi_2| \\
\lambda_{21}|\varphi_2\rangle \langle \varphi_1| \qquad &
\lambda_{22}|\varphi_2\rangle \langle \varphi_2|
\end{array} \right ),
\label{mod5}
\end{equation}
with $\lambda_{12}=\lambda_{21}$. This form corresponds
basically to the general separable potential derived in Eq.(\ref{sp}). For
all form factors entering $V_{NqN}$  we use the
simple Gaussian form with one scale parameter $r_0$:
$\varphi_i(r)=
 Nr^{L_i+1} \exp (- r^2/2r_0^2)$.
 In the calculations   the averaged value of pion-nucleon
coupling constant  $f^2_{\pi NN}/(4\pi ) = 0.075$ and a soft cut-off
parameter with  values $\Lambda_{\pi NN}=\Lambda_{\rm dip} = 0.60\div 0.73 $~GeV
have been used. The results of the fits of the model parameters $\lambda_k $
(or $\lambda_{jk}$), $r_0$ and $E_0$ to the $NN$ phase shift analysis data
are displayed on Fig.~2\footnote{The respective values of parameters can be
found in \cite{Ku2000}.}. The parameter $E_0$ corresponds to the sum of the
six-quark bag energy and the effective $\sigma$-meson mass inside the
six-quark bag. It is quite evident this  simple model is able to describe
the $NN$ low partial waves up to   $E_{\rm lab}=600$~MeV very well. The
respective phase shifts and the mixing parameter $\varepsilon_1$ are
compared in Fig.~2 with data of a recent phase shift analysis (SAID,
solution SP99).

 Moreover, it was very  surprising to find out that such a simple model gives
very good description   for $^1S_0$ phase shifts even up to $E_{\rm
lab}=1200$~MeV~\cite{Ku2000}. It is very instructive here to discuss
the description  of phase shifts in
triplet coupled channels $^3S_1 -  {}^3D_1$, and especially the
behavior of the mixing parameter  $\varepsilon_1$  with increasing energy.
Without the (quark-bag induced) non-diagonal mixing potential (i.e. at
$\lambda_{12}=0$) the behavior of $\varepsilon_1$ is correct only at very
low energies (see the dashed line on the lower right panel
in Fig.~2).  The increase of the
cut-off parameter $\Lambda_{\pi NN}$ up to values 0.8 GeV does not help to
get a better agreement with the data, but on the contrary, destroys  the
good description at low  energies (the dotted line in the panel for
$\varepsilon_1$ in Fig.~2).
Introducing the quark-bag induced mixing ($\lambda_{12}\ne  0 $ in
Eq.(\ref{mod5})), which we predict on the base of the suggested new
mechanism, allows us to reproduce  the behavior of  $\varepsilon_1$ (and
$^3S_1 - {}^3D_1$ phase shifts as well) with a reasonable accuracy until the
energy as high as $E_{\rm lab}\sim 600$~MeV, but for sufficiently small
values of $\Lambda_{\pi  NN}$  only. The best fit result for the $\varepsilon_1$
mixing parameter is shown on the lower right panel of Fig.~2 (by solid line)
and is attained with the value of $\Lambda_{\pi NN}=0.594$~GeV.
In this case the condition  $\lambda_{12}^2=\lambda_{11}\lambda_{22}$
is satisfied with high accuracy. Just this  condition follows from  our
assumption that the quark-bag induced  $S-D$ mixing arises due to coupling
of the $NN$ channel with $L=0,2$ to a single  $S$-wave six-quark state
$|s^6+2\pi \rangle$ (see Eq.(\ref{sp})). The  increase of $\Lambda_{\pi NN}$
up to a value 0.8~GeV (with keeping the best fit to the phase shifts)
results in the violation of the above condition and in a significant
deterioration of the description of $\varepsilon_1$ (the dot-dashed curve in
the lower right panel of Fig.~2).

Another important result of the present model could be a possible
resolution of a long-standing puzzle about the small vector-meson
contribution  in single-baryon spectra and a strong spin-orbital splitting
(due to
the vector  meson contribution) in the $NN$ interaction.  Our
explanation of  the puzzle is  based on
the fact that there is no significant  vector-meson  contribution to $qq$
force (in $t$-channel) but there is an important
contribution  of vector mesons in dressing the symmetric  six-quark
bag  leading thereby to strong spin-orbital effects in the $NN$  interaction
mediated by the "dressed" bag.
 Moreover, the proposed model  will lead to the appearance of strong $3N$
and $4N$ forces  mediated by $2\pi$ and $\rho$ exchanges \cite{Kuk99,Ku2000}.
The  new $3N$ forces include both central and spin-orbit components.
Such a spin-orbit $3N$ force is extremely desirable to explain the low
energy puzzle of the analyzing power  $A_y$ in $N$-$d$
scattering and also the behavior of $A_y$ in the $3N$
system at higher energies $ E_N\simeq 250\div 350 $~MeV at backward
angles~\cite{Fost99}. The  central components of the $3N$ and $4N$ forces are
expected to be  strongly attractive  and thus they must contribute to $3N$
and (may be) $4N$ binding energies possibly resolving hereby the very old puzzle
with the binding energies of the lightest nuclei.

To conclude:
In this work we presented a new mechanism for the description of the
intermediate- and short-range $NN$ interaction.
The  mechanism is distinguished from the  traditional Yukawa concept of meson
exchange in the $t$-channel. Instead of this, we introduce here
a concept of the dressed  symmetric six-quark bag in  the intermediate state
with $s$-channel  propagation. The new interaction mechanism proposed
here  has been shown to lead to separable energy-dependent $s$-channel
interaction with nodal form factors which reflect the orthogonality of
six-quark configurations in the initial $NN$
 channel and the intermediate bag-like state.

Using a simple illustrative model we found that  it is possible to give a
good description of
 all the lowest $NN$ phase shifts  in a large energy interval $0 \div 600$~MeV.
 This suggests strongly that the new microscopic mechanism of $s$-channel
"dressed" symmetric bag should work adequately.
 The model gives a natural
microscopic background for previous semi-phenomenological models
like the Moscow $NN$  potential, the Tabakin separable potential "with
attraction and  repulsion" and also the QCB model by
Simonov and other hybrid  models.
The significant enhancement of $\sigma$- and $\rho$-meson fields around the
symmetric six-quark bag may also contributes to an explanation of ABC
puzzle alternative to that assumed today. Future studies  will show
to what degree such expectations can be  justified.

\centerline{\large \bf Acknowledgments.}  The authors express their deep
gratitude to
Profs. Walter Gloeckle, Amand Faessler, Steven  Moszkowski and Mitja Rosina
for the fruitful discussions.  The
 authors  thank the  Russian Foundation for Basic Research (grant
RFBR-DFG No.92-02-04020) and the  Deutsche Forschungsgemeinschaft
(grant No.  Fa-67/20-1) for partial financial  support.


\centerline{\large \bf Figure captions}

\bigskip

{\bf Fig.~1.} The graph illustrates the mechanism of $NN$ interaction via 
two sequential $\pi$-meson emissions and
absorptions via an intermediate $\sigma$- (or $\rho$-) meson and the
generation of a dressed six-quark bag $DB=s^6+\sigma(\rho)$ in $NN$
scattering.

\bigskip
{\bf Fig.~2.}  The $NN$ phase shifts (in deg.) as predicted by our
illustrative  model in comparison with PSA data (SAID,  solution SP99).
The mixing parameter $\varepsilon_1$ for different
values of cut-off parameter $\Lambda_{\pi NN}$ is displayed in the lower
right corner (for explanations see the text).

\newpage

\begin{figure}[hp]\centering
\epsfig{file=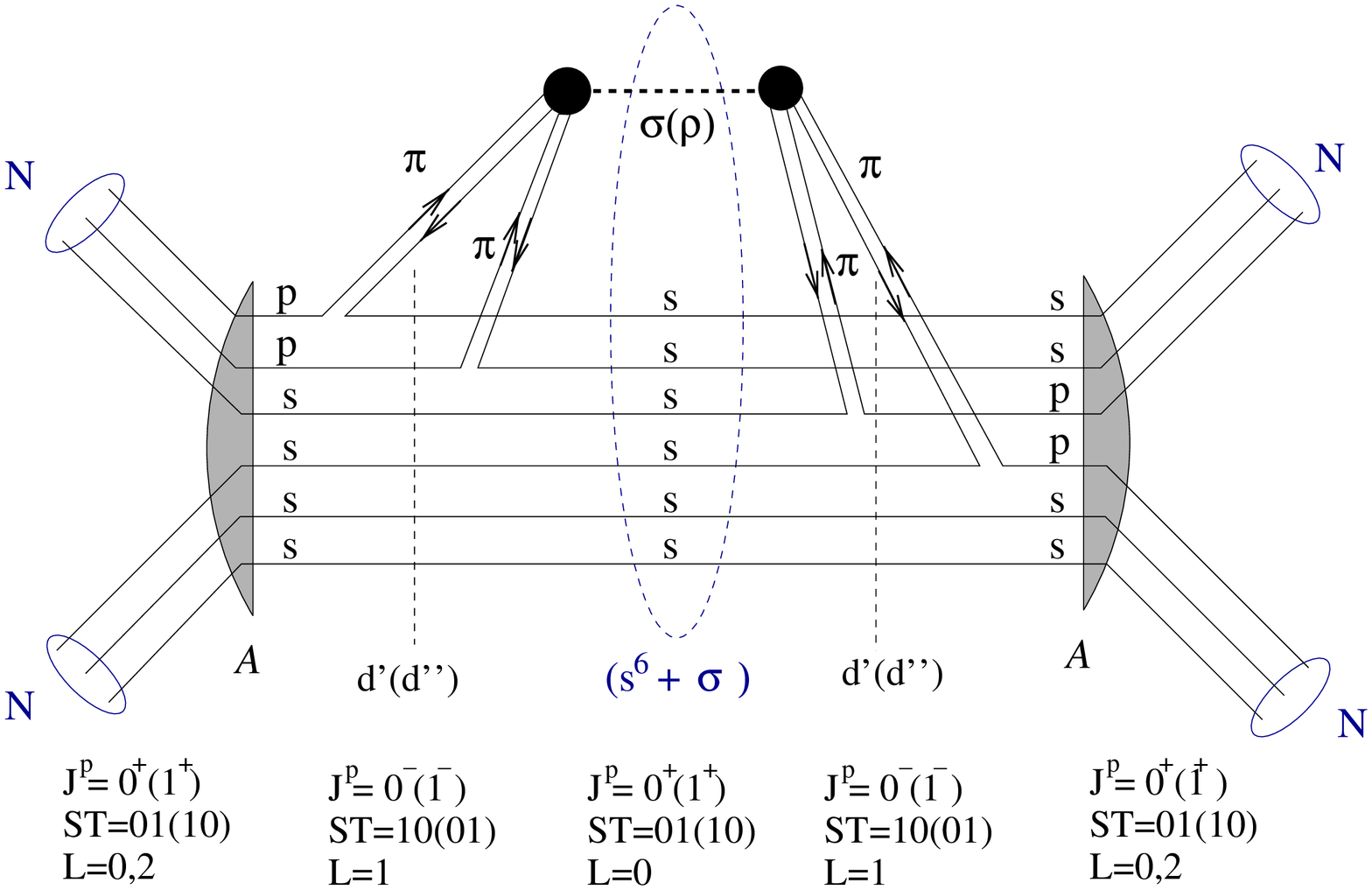,height=0.6\textheight,width=0.5\textwidth,angle=-90}
\vspace{0.5cm}
 \caption{\small  }
\end{figure}

\vspace{1.5cm}

\begin{figure}[hp]\centering
\epsfig{file=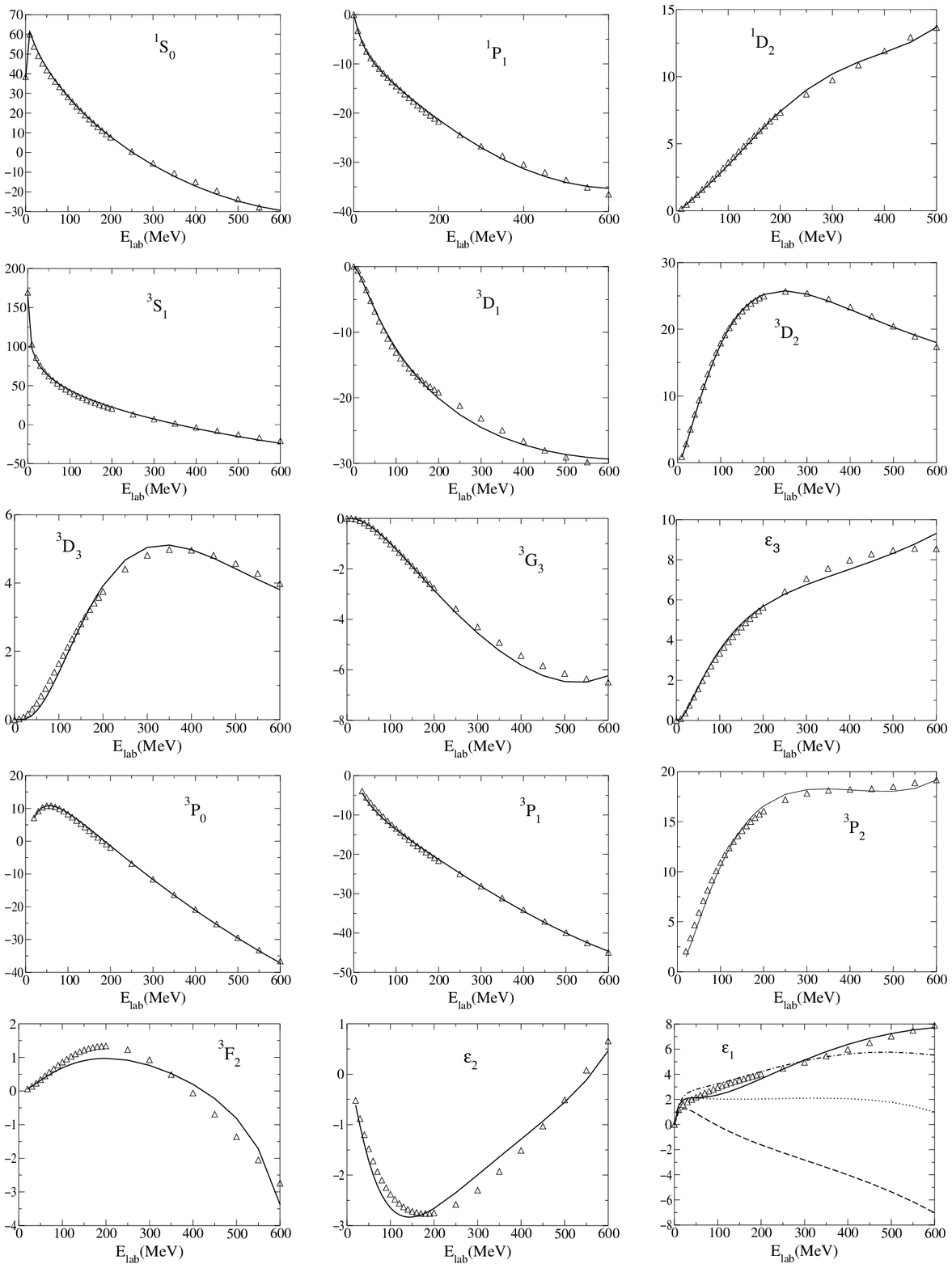,height=0.95\textheight}
 \caption{\small }
\end{figure}


\begin{thebibliography}{99}
\bibitem{Mach87} R. Machleidt, K. Holinde and Ch. Elster, Phys.~Rep. {\bf 149},
 1 (1987);
 R. Machleidt, Adv.~Nucl.~Phys. {\bf 19}, 189 (1989).
\bibitem{Kuk99} V. I. Kukulin, {\em Proceeds. of the V Winter School on
 Theoretical Physics PIYaF, Gatchina, S.-Petersburg, 8 -- 14 Febr. 1999.},
 p.~142.

\bibitem{FB98} {\em Proceeds. of 16th European Conf. on Few-Body Problems in Physics, 1998,
 Autrans}, Suppl. of Few-Body~Systems, {\bf X} (1999).
\bibitem{Fost99} R. D. Foster et al., TUNL {\bf 37}, Progress Report, p.31,
 1999;
 G. Martinus, O. Scholten and J. Tjon. Phys. Lett {\bf B
 402}, 7 (1997); Phys. Rev. {\bf C 50} (1997), 2945 (1997); {\em ibid} {\bf C
 58}, 686 (1997);
 H. O. Meyer, {\em A talk at Symposium on Current Topics in the
Field of Light Nuclei, Cracow, Poland, 21 -- 25 June 1999}, p.~858,
(unpublished).
\bibitem{Pla94} D. Plaemper, J. Flender and M. F. Gari, Phys.~Rev.
  {\bf C 49}, 2370 (1994);
 T.-S. H. Lee, Phys.~Rev. {\bf C 29}, 195 (1984);
      T.-S.~H.~Lee and A.~Matsuyama, Phys.~Rev. {\bf C 32}, 516 (1985);
      A.~Matsuyama and T.-S.~H.~Lee, Phys.~Rev. {\bf C 34}, 1900 (1986).
\bibitem{Kus91} A. M. Kusainov, V. G. Neudatchin, and I. T. Obukhovsky,
 Phys.~Rev.~C {\bf 44}, 2343 (1991); I. T. Obukhovsky, Prog.~Part.~Nucl.~Phys. {\bf 36}, 359
(1996).
\bibitem{PRC99} V. I. Kukulin, V. N. Pomerantsev,  and A. Faessler,
    Phys. Rev. {\bf C 59}, 3021 (1999);
    V. I. Kukulin and V. N. Pomerantsev, Progr. Theor. Phys. {\bf 88}, 159
    (1992).
\bibitem{Kun94} T. Hatsuda and T. Kunihiro, Phys.~Rep. {\bf 247}, 221 (1994);
 T.~Hatsuda, T.~Kunihiro and H.~Shimizu, Phys.~Rev.~Lett. {\bf 82},
 2840 (1999); P. Rehberg, L. Bot and J. Aichelin, Nucl. Phys. {\bf A653}, 415
(1999).
\bibitem{Ku2000} A. Faessler, V. I. Kukulin, I. T. Obukhovsky and V.~N.~Pomerantsev,
    E-print:nucl-th/9912074.
\bibitem{Stancu} D. Barz and Fl. Stancu, Phys.~Rev. {\bf C 60}, 055207
(1999).
\bibitem{Obu99} I. T. Obukhovsky, Amand Faessler, Georg Wagner and  
A. J. Buchmann, Phys. Rev. C {\bf 60}, 035207 (1999). 
\bibitem{Micu69} L. Micu, Nucl. Phys. {\bf B10}, 521 (1969);
  A.~Le~Yaouanc, L.~Oliver, O.~Pene and J.~C.~Raynal, Phys.~Rev.~D {\bf 8},
     2223 (1973); {\bf 9}, 1415 (1974); {\bf 11}, 1272 (1974).
\bibitem{Guts94} T. Gutsche, R. D. Viollier and A. Faessler, Phys. Lett. {\bf
331}, 8 (1994).
\bibitem{Harv81} M. Harvey, Nucl. Phys. {\bf A352}, 301 (1981);
  {\bf A352},  326 (1981).
\bibitem{KukMac} V. I. Kukulin, V. N. Pomerantsev, S. G. Cooper and R.~Mackintosh,
 Few-Body Systems, Suppl. {\bf 10}, 439 (1998).

\end{thebibliography}
\end{document}